\documentclass[10pt]{article}

\usepackage[utf8]{inputenc} 
\usepackage[T1]{fontenc}    
\usepackage{amsmath}        
\usepackage{amssymb}        
\usepackage{graphicx}       
\usepackage{hyperref}       
\usepackage{enumitem}       
\usepackage{geometry}       
\usepackage{authblk}        
\usepackage{pgfplots} 
\usepackage{multirow} 
\usepackage{tikz} 
\usepackage{array} 
\usepackage{nicefrac}
\usepackage[backend=biber, style=numeric]{biblatex}
\usepackage{orcidlink}
\usepgfplotslibrary{groupplots} 
\pgfplotsset{compat=1.18} 
\title{Solving tricky quantum optics problems with assistance from (artificial) intelligence} 

\author[1]{Manas Pandey \orcidlink{0009-0000-4409-940X}} 
\author[2]{Bharath Hebbe Madhusudhana \orcidlink{0000-0001-8482-1778}} 
\author[1]{Saikat Ghosh \orcidlink{0000-0002-2988-3789}}
\author[3,4,5,6]{Dmitry Budker \orcidlink{0000-0002-7356-4814}} 

\affil[1]{Indian Institute of Technology, Kanpur, Uttar Pradesh 208016, India}
\affil[2]{MPA-Quantum, Los Alamos National Laboratory, Los Alamos, NM 87544, United States}
\affil[3]{Johannes Gutenberg-Universit{\"a}t Mainz, 55122 Mainz, Germany}
\affil[4]{Helmholtz Institute Mainz, 55099 Mainz, Germany}
\affil[5]{GSI Helmholtzzentrum für Schwerionenforschung GmbH, 64291 Darmstadt, Germany}
\affil[6]{Department of Physics, University of California, Berkeley, CA 94720-7300, United States of America}

\date{\today} 

\begin{document}

\maketitle 
\begin{abstract}
The capabilities of modern artificial intelligence (AI) as a ``scientific collaborator'' are explored by engaging it with three nuanced problems in quantum optics: state populations in optical pumping, resonant transitions between decaying states (the Burshtein effect), and degenerate mirrorless lasing. Through iterative dialogue, the authors observe that AI models—when prompted and corrected—can reason through complex scenarios, refine their answers, and provide expert-level guidance, closely resembling the interaction with an adept colleague. The findings highlight that AI democratizes access to sophisticated modeling and analysis, shifting the focus in scientific practice from technical mastery to the generation and testing of ideas, and reducing the time for completing research tasks from days to minutes.
\end{abstract}
\newpage

\section{Introduction}
\label{sec:introduction}

The rapid emergence of artificial intelligence (AI) is changing the way science is done. As with many new tools (calculators, e-mail, internet, etc.), we usually begin by applying these tools to solve common tasks better than it is possible with existing tools (e.g., performing arithmetical operations with a calculator rather than a slide rule). However, the real power of new tools lies in enabling completely new uses such as collaborative paper writing with colleagues anywhere in the world, enabled by the Internet. We are convinced that the use of AI in science will bring a plethora of new uses and capabilities, some of which are already apparent today. An example is that AI is ``democratizing'' science by enabling any reasonably qualified scientist to perform sophisticated modeling using highly specialized algorithms, without the need to master software packages.  AI takes the role of an expert colleague who can understand the ``professor'' formulating the question and is also able to run the dedicated software, thus eliminating the ``middleman''.

Here, we describe how we test AI abilities and new ways of ``interacting with the tool'' with three problems in quantum optics:

\begin{enumerate}[label=\roman*., itemindent=*, leftmargin=3em] 
    \item A straightforward question; however, known to trick even mature physicists in the field.
    \item A subtle problem with important applications that, while known for some years, is still a subject of current research.
    \item A problem of current research with an unsettled solution.
\end{enumerate}

Based on experience with these problems, we make observations regarding the possible utility of modern AI in the scientific process. In essence, every scientist now has access to sophisticated tools previously accessible only to specialists. This brings forward the importance of ideas rather than techniques. Of course, the speed with which ideas can be elaborated, tested in detail, and perhaps executed is higher. What used to take months and years can now be done in minutes. 

We also remark on the striking similarity of the AI behavior to that of students. 

We need to explain here what exactly we mean by ``AI''. We test our problems on various state-of-the-art general-purpose models, accessible for public use (e.g., Gemini 2.5\,Pro). When using different models, the details of the dialog are different; however, the overall results are broadly consistent across AI platforms. Therefore, unless we are specifically interested in comparing the models, we generically describe the interactions with AI in the following. The dialogs with the AI on which this paper is based can be found on a dedicated web site \url{https://manasp21.github.io/AI_QO_Appendix/index.html}.


\section{Problem One: State populations upon optical pumping}
\label{sec:problem_one}

The first problem we discuss is a rather straightforward quantum optics question. However, in our experience, most physicists, including experts in the specific subfield, tend to give an incorrect answer, at least initially. 

\begin{figure}
    \centering
    \includegraphics[width=0.5\linewidth]{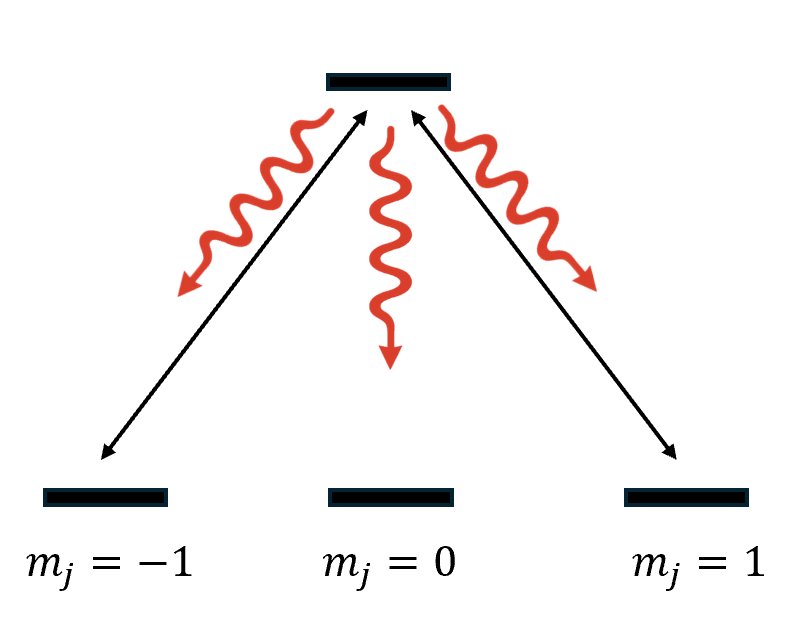}
    \caption{\textbf{Problem One:} A $J=1\rightarrow J'=0$ transition optically pumped with $x$-linearly polarized light. Spontaneous decay from the excited state is the only relaxation process considered in the problem.}
    \label{fig1}
\end{figure}

\subsection{Formulation of the problem}
\label{subsec:problem_one_formulation}

Consider an ensemble of stationary atoms in the ground $J=1$ state (see Fig.\,\ref{fig1}), in the absence of external fields. The atoms are optically pumped with monochromatic light propagating along $z$, which is linearly polarized along $x$. The excited state has angular momentum $J=0$ and it decays spontaneously back to the ground state. There is no relaxation in the system other than this spontaneous decay. Initially, the ground state is unpolarized, so that the population is equally distributed among the ground-state sublevels.

Assuming that the initial population of each of the ground-state Zeeman sublevels is 1/3, what are the populations after optical pumping is complete?

\subsection{``Human'' solution}
Most physicists presented with this problem initially think that the populations of the $m_j=-1$, 0, and 1 sublevels are 0, 1, and 0, respectively (see Table\,\ref{tab:1}, the first line in ``Final population''). Indeed, atoms in the $m_j=0$  cannot be excited by light, in contrast to the atoms in the other two sublevels. If an atom excited from a $|m_j|=1$ sublevel spontaneously decays back into one of these two sublevels, it has a chance to get reexcited, until it lands in the ``dark'' $m_j=0$ state.

What is wrong with this solution is that it ignores the fact that there is not one but two dark states in the problem. The second dark state is an equal-weight superposition of $m_j=-1$ and $m_j=1$, which is
\begin{equation}
\frac{1}{\sqrt{2}}     \left(|m_j=-1\rangle + |m_j=1\rangle\right)
\label{Eq:dark_state}
\end{equation}
for $x$-polarized light. 

One way to see that there are indeed two dark states is to look at the same problem with the quantization axis chosen along the light polarization. In this case, light only drives the $\Delta m_j=0$ transition, and the $|m_j|=1$ sublevels are both ``dark''.

Upon realization that only atoms in the ``bright'' superposition of the $|m_j|=1$ sublevels orthogonal to the superposition \eqref{Eq:dark_state}, some physicists conclude that the final populations are 1/6, 2/3 and 1/6. However, this is wrong again as this ignores the spontaneous decay of the excited state to the dark superposition.

Finally, the correct solution for the populations is 1/4, 1/2, and 1/4, which can be verified, once again, by turning to the picture with the quantization axis chosen along the light polarization: the final population of each of the dark states is equal to 1/2 of the total initial population and the sole dark state is empty.

\renewcommand{\arraystretch}{1.5}

\begin{table}
    \centering
    \begin{tabular}{|l|>{\centering\arraybackslash}m{2cm}|>{\centering\arraybackslash}m{2cm}|>{\centering\arraybackslash}m{2cm}|}
    \hline
    State & $m_J = -1$ & $m_J = 0$ & $m_J = 1$ \\
    \hline
    Initial population & $\nicefrac{1}{3}$ & $\nicefrac{1}{3}$ & $\nicefrac{1}{3}$ \\
    \hline
    \multirow{3}{*}{Final population} & 0 & 1 & 0 \\
    \cline{2-4}
    & \nicefrac{1}{6} & \nicefrac{2}{3} & \nicefrac{1}{6} \\
    \cline{2-4}
    & \nicefrac{1}{4} & \nicefrac{1}{2} & \nicefrac{1}{4} \\
    \hline

\end{tabular}
    \caption{\textbf{Problem One:} A summary of the sublevel populations . The ``Final population'' section summarizes some variants of the answer. While all of them are reasonable ``educated guesses'', only the last one is correct.}
    \label{tab:1}
\end{table}

\subsection{Interaction with AI}
\label{subsec:problem_one_interaction}
We posed the problem to the AI and it immediately recognized that this is a question in the field of quantum optics. It proceeded to give us the 0,1,0 answer (see Table.\,\ref{tab:1}). While this answer is incorrect, it can be considered a reasonable first attempt. If a student gives such an answer and can motivate it, he or she will definitely ``earn points'' for this. 

In the next step, we prompted the AI that we will try to explain that there is a flaw in the solution by solving a related problem, specifically the same problem but with the quantization axis along the light polarization. The AI solved this correctly. We then asked to write the density matrix and rotate it to the basis of the original problem, pointing out that this density matrix is different from the one originally presented by the AI. 

The AI proceeded to acknowledge that it had been wrong and wrote: ``Thank you! This detailed comparison has been very illuminating and has helped correct a flaw in one of my earlier consistency check analyses.''

We see that the interaction with the AI is quite similar to that with an excellent student or knowledgeable and intelligent colleague.

\section{Problem Two: Resonant transitions between decaying states (the Burshtein effect)}
\label{sec:problem_two}

\subsection{Formulation of the problem}
\begin{figure}
    \centering
    \includegraphics[width=0.99\linewidth]{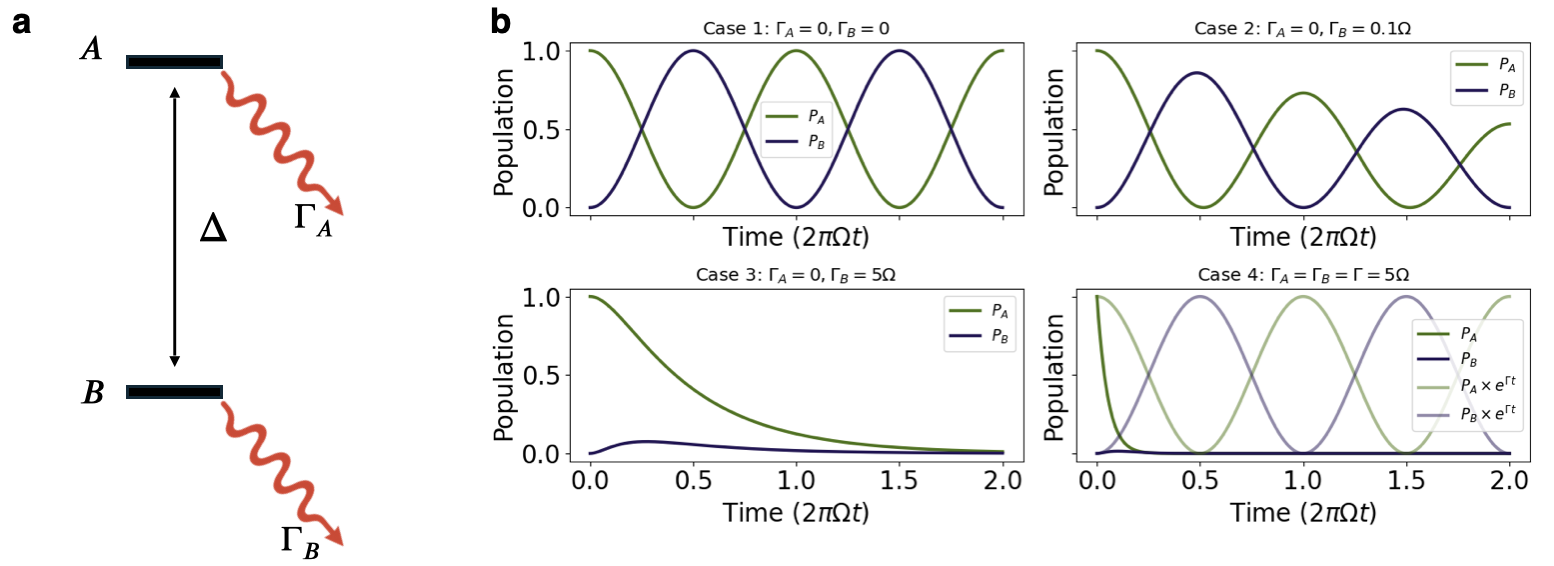}
    \caption{\textbf{Problem Two:} \textbf{a.} Two decaying states coupled resonantly. \textbf{b.} The populations of states $A$ and $B$ for the four cases described in the text. Time is presented as a dimensionless quantity (number of Rabi cycles). }
    \label{fig2}
\end{figure}

Consider two states, $A$ and $B$, with energy gap $\Delta$ between them (Fig.~\ref{fig2}a). At time t=0, state $A$ is populated and state $B$ is empty. A resonant excitation field drives the transitions between the two states. The strength of the driving field is characterized by the Rabi frequency $\Omega$. We are interested in the population of state $B$ as a function of time for several different cases for the relaxation rates of the two states $\Gamma_A$ and $\Gamma_B$; it is assumed that the states decay into unobserved states:
\begin{enumerate}[label=Case\,\arabic*: , leftmargin=6em]
    \item $\Gamma_A=\Gamma_B=0$\,;
    \item $\Gamma_A=0;\,\,\Gamma_B\ll \Omega$\,;
    \item $\Gamma_A=0;\,\,\Gamma_B\gg \Omega$\,;
    \item $\Gamma_A=\Gamma_B=\Gamma\gg \Omega$\,.
\end{enumerate}

\subsection{``Human'' solution}
Up to a point, this problem is a standard textbook problem of a two-level system in the presence or a resonant periodic perturbation (see, for example, \cite{OptPolAt}, Sec.\,7.1). When the relaxation rates of states A and B are both zero (case 1), we have Rabi oscillations of the population between the two states [Fig.\,\ref{fig2}b, top left]. Adding a small decay for one of the states (case 2), the oscillations decay exponentially [Fig.\,\ref{fig2}b, top right]. When the decay rate becomes large compared to the Rabi frequency, the system is in the overdamped regime [Fig.\,\ref{fig2}b, bottom left] and there are no oscillations seen. We note that the population of $B$ is zero at $t=0,\infty$ and never at intermediate times.  Finally, we consider the most interesting case 4 [Fig.\,\ref{fig2}b, bottom right]. Again, as in case 3, the populations decay. However, and importantly, the solution remains oscillatory despite the rapid decay. To illustrate this, we multiply 
the solution by a growing exponential $e^{\Gamma t}$. We see that this operation completely restores the solution to that of case 1. In particular, the population of $B$ turns to zero at the same periodicity as for Rabi oscillations in case 1 (no decay). This phenomenon was pointed out by A.I.\,Burshtein and coworkers in the 1980s \cite{Burshtein1988}. 

The ``Burshtein effect'' has important implications for metrology. If one measures the time evolution with excellent signal-to-noise and the states decay at the same rate, one can compensate for the exponential decay and obtain full information about the coupling of states $A$ and $B$,  which is impossible when the decay rates of $A$ and $B$ are vastly different (for example, as in case 3).   

\subsection{Interaction with AI}

\label{subsec:problem_two_interaction}

When we presented the problem of the two-level system with its varying decay scenarios to the AI, Gemini, it promptly recognized the context as a foundational one within quantum optics. The AI then proceeded to address the calculation of the population of state $B$, by systematically tackling each of the four enumerated conditions.
Its methodology for each case appeared to be:
\begin{enumerate}
    \item \textbf{Framing the Physics:} The AI initially framed the problem in terms of coherent Rabi driving ($\Omega$) counteracted by the specified incoherent decay channels ($\Gamma_A, \Gamma_B$).
    
    \item \textbf{Systematic Case Analysis:}
    \begin{itemize}
        \item For \textbf{Case 1 ($\Gamma_A=\Gamma_B=0$)}, it provided the standard $\sin^2(\Omega t/2)$ solution, characteristic of undamped Rabi oscillations.
        \item In \textbf{Case 2 ($\Gamma_A=0; \Gamma_B\ll \Omega$)}, its solution appropriately introduced an exponential decay factor enveloping the Rabi oscillations, reflecting the impact of the slow decay of state $B$.
        \item For \textbf{Case 3 ($\Gamma_A=0; \Gamma_B\gg \Omega$)}, its analysis correctly shifted to an overdamped perspective. It highlighted how the rapid decay of the state $B$ suppresses the oscillations and severely limits the maximum population transferred. 
        \item In \textbf{Case 4 ($\Gamma_A=\Gamma_B=\Gamma\gg \Omega$)}, the AI described a scenario where strong decay in both states heavily dampens the coherent evolution, leading to only a minor, transient population in state $B$. \textbf{While the statements given by the AI thus far are reasonably correct, it failed to recognize the Burshtein effect.}
    \end{itemize}
\end{enumerate}
At this point, we decided to proceed as in the case of Problem One and ``convince" the AI to re-examine and refine its solution. However, while we were doing this, an updated AI model was released, and it handled the problem perfectly from the first attempt, including an excellent discussion of the Rabi oscillation underneath the exponential decay envelope.


\section{Problem Three: degenerate mirrorless lasing}
\label{sec:problem_three}
\subsection{Formulation of the problem}
\label{subsec:problem_three_formulation}
Mirrorless lasing (MR) is a process of generation of a directed nearly monochromatic radiation, without any mirrors, by a medium (in our case, an ensemble of atoms) exposed to a pump laser beam. ML is usually associated with ``pencil-like'' pumping; it often occurs in both forward and backward direction with respect to the pump beam. Degenerate mirrorless lasing (DML) refers to a situation where the emitted radiation is at a frequency that is the same (or close to) the frequency of the pump light. 

\begin{figure}
    \centering
    \includegraphics[width=0.99\linewidth]{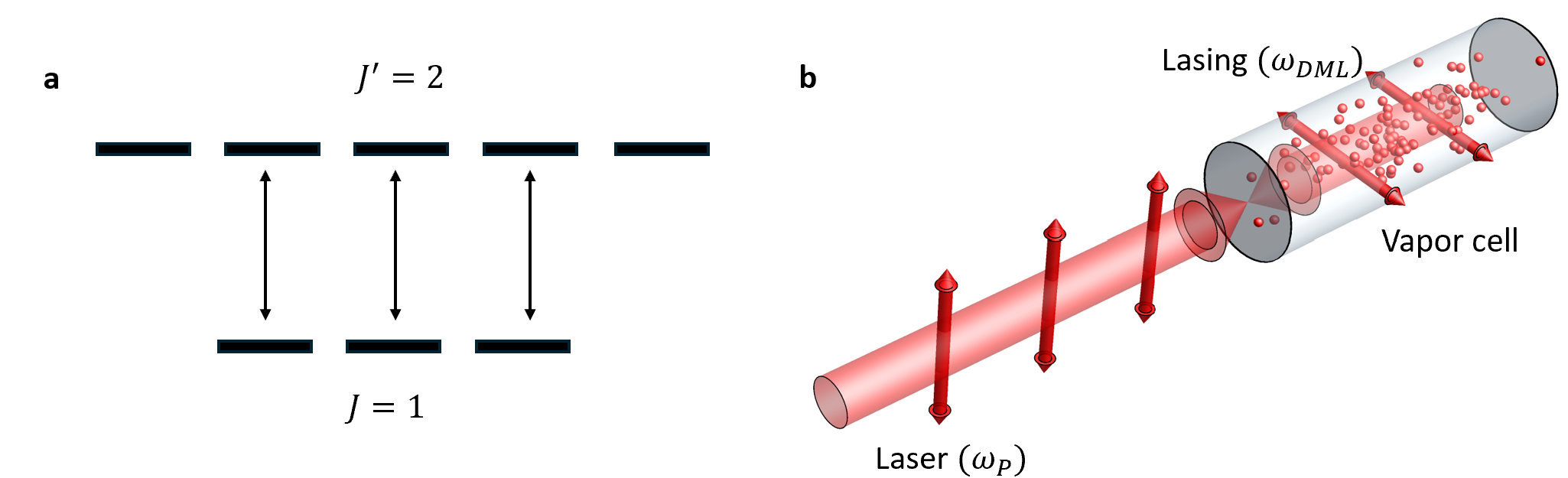}
    \caption{The setup for a backward-DML: a) The energy levels and transitions along the lines of the original proposals \cite{gazaryan_grigoryan_papoyan_2011,movsisyan2011amplification}. b) Experimental arrangement.}
    \label{fig3}
\end{figure}

Consider a $J=1 \rightarrow J'=2$ transition [Fig.\,\ref{fig3}a]. There is no relaxation in the system except for the spontaneous decay from the upper state to the lower state. The atoms are stationary. The pumped volume is a cylinder of length $L$ and radius $R$; the atomic number density is $n$. Atoms are illuminated with resonant cw light with linear $z$ polarization. 

What are the conditions for DML in the forward and backward directions for light with $x$ polarization? The axis of the cylinder is along the $y$ axis.
\subsection{``Human'' solution}
As mentioned in the Introduction, this is actually an unsolved problem. Reference \cite{ramaswamy2023mirrorless} reports that backwards DML was observed in experiments in Ashtarak; however, these results could not be reproduced in Mainz. The most recent theoretical analysis \cite{ramaswamy2023mirrorless} points out that the simplified theoretical treatment in the original proposal papers \cite{gazaryan_grigoryan_papoyan_2011,movsisyan2011amplification} is insufficient in the high-light-power conditions that would be needed to achieve the desired lasing in a thermal vapor. A recent theoretical suggestion \cite{ramaswamy2025degenerate} to use large detunings is awaiting experimental verification.  
\textbf{}
\subsection{Interaction with AI}
Given the formulation of the problem as a prompt, the AI immediately correctly identified the area of physics and found relevant literature. Its comments on the topic were somewhat general, however, certainly useful for somebody not yet intimately familiar with the field. We noticed that it would not commit to a specific answer to the problem, however, it would make statements along the lines of ``the backwards DML may occur under the right conditions'', going on to specify that the atomic density, pump laser power, geometry, etc. are the important factors.  At this point, we asked AI the following: ``I am a researcher and I want to set up an experiment to detect backwards DML in a thermal rubidium vapor, can you help me design the parameters of the experiment?'' The response was a detailed and substantive discussion, at a high professional level, of the design considerations, how one should choose the parameters, and what aspects one needs to pay particular attention to. We felt that this kind of advice was similar to what, say, a PhD student would get from a highly knowledgeable and experienced colleague. We remarked among ourselves that we should, indeed, recommend students to consult AI when preparing experiments and that this represents a powerful tool that was not available in pre-AI times (except for rare cases when a suitably knowledgeable, experienced and patient human colleague could be found). Finally, we asked AI to suggest the reasons that backwards DML could to be observed in one lab but could not be reproduced in another. Again, the response was a thoughtful, professional answer listing a number of possible explanations. However, for this latter question, it is hard to delineate whether the AI used its direct knowledge of the discussion in the literature or came up with the idea based on an original thought process. The ``show the thinking process'' feature of the AI model indicates that the model has not explicitly used experimental papers to come up with its answers. 

\label{subsec:problem_three_interaction}


\section{Discussion}
\label{sec:discussion}
Our experience as researchers in the field using AI to help navigate nontrivial questions in quantum optics, exemplified by the three problems above and many other similar experiences, convinces us that we have an unprecedented, powerful tool at our disposal that has already changed the way we research a scientific topic, analyze discrepancies, and design new experiments.

As with any new tool, one needs to learn to use it properly. While some people tend to put full trust into AI, it is, in fact, not an omniscient being, and the answers it gives are often wrong. The art of using AI is in taking full advantage of it despite this. We conclude that the old human wisdom pointing to the importance of asking the right questions equally applies to interactions with the AI. We derive the most benefit by treating the AI as a multidisciplinary collaborator and not as a calculator or a handbook with ready answers.

Scientists have a toolbox for checking physics solutions. These include checking the units and limiting cases, solving similar problems where the solutions are known, solving the same problem in multiple ways, checking that the solution possesses the required symmetries (e.g., basis invariance as on our Problem One), satisfies conservation laws, etc. Not only can AI be used to perform these tests, it can be prompted (read: \emph{taught}) to check its own answers as well as not to make the same type of a mistake twice. Our experiences show that this approach can help AI overcome its own ``biases''. 






\section{Conclusions}
\label{sec:conclusions}
We found AI to be an extremely helpful, efficient ``colleague'', while not immune to errors and misconceptions, however, eagerly ready to efficiently deal with them in cooperation with us, the ``user". The speed with which such interaction progresses is already incomparably faster than purely human interaction, and this is not exclusively due to the availability of knowledge on the internet. The speed of the intellectual synthesis is also increased significantly. While we have not attempted a scientific quantification, we feel that a discussion that would normally take days can be compressed to less than an hour.

A new age in science has begun.

\section*{Acknowledgements}
\label{sec:acknowledgements} 

We thank Dr.\,Anahit Gogyan, Prof. Stefan Kramer, Marina Gil Sendra, Andrew Budker, Prof. Derek Jackson Kimball, Prof. Alexej Jerschow, Prof. Ilya Kuprov, Prof. Mikhail Kozlov, Prof. Piet O. Schmidt, Prof. Ferdinand Schmidt-Kaler, Prof. Svetlana Malinovskaya, Dr. Aneesh Ramaswami, Dr. Aleksandra Ziolkowska, Erin Burgard, Dr. Andreas Trabesinger, and others for helpful discussions. Research presented in this article was supported in part by the Laboratory Directed Research and Development program of Los Alamos National Laboratory under project number 20230779PRD1.  

\label{sec:references}
\printbibliography


\appendix 
\end{document}